\documentclass[useAMS,usenatbib]{mn2e}
\usepackage{url}
\usepackage{graphicx}
\usepackage{verbatim}
\usepackage[usenames]{color}
\DeclareGraphicsExtensions{.pdf,.png,.jpg,.mps,.eps,.ps}
\usepackage{amsmath}
\usepackage{natbib}
\usepackage[flushleft]{threeparttable}

\usepackage{ulem}

\newcommand\suzaku{{\it Suzaku}}

\newcommand\rxte{{\it RXTE}}
\newcommand\xmm{{\it XMM-Newton}}

\newcommand\s{{\rm~s}}
\newcommand\ks{{\rm~ks}}

\newcommand\kpc{{\rm~kpc}}

\newcommand\hz{{\rm~Hz}}
\newcommand\kev{{\rm~keV}}

\newcommand\kms{\ifmmode {\rm~km\ s}$^{-1}$ \else ~km s$^{-1}$\fi}
\newcommand\Hunit{\ifmmode {\rm~km\ s}$^{-1}$\ {\rm Mpc}$^{-1}$
        \else ~km s$^{-1}$ Mpc$^{-1}$\fi}
\newcommand\ctssec{\ifmmode {\rm~count\ s}$^{-1}$ \else ~count s$^{-1}$\fi}
\newcommand\ergsec{\ifmmode {\rm~erg\ s}$^{-1}$ \else
        ~erg s$^{-1}$\fi}
\newcommand\funit{\ifmmode {\rm~erg\ s}$^{-1}$\ ; {\rm cm}$^{-2}$ \else
        ~ergs s$^{-1}$ cm$^{-2}$\fi}
\newcommand\phflux{\ifmmode {\rm~photon\ s}$^{-1}$\  ; {\rm cm}$^{-2}$
        \else   ~photon s$^{-1}$ cm$^{-2}$\fi}
\newcommand\efluxA{\ifmmode {\rm~erg\ s}$^{-1}$\ ; {\rm cm}$^{-2}$\ ; {\rm
        \AA}$^{-1}$ \else ~erg s$^{-1}$ cm$^{-2}$ \AA$^{-1}$\fi}
\newcommand\efluxHz{\ifmmode {\rm~erg\ s}$^{-1}$\ ; {\rm cm}$^{-2}$\ ; {\rm
        Hz}$^{-1}$ \else ~erg s$^{-1}$ cm$^{-2}$ Hz$^{-1}$\fi}
\newcommand\cc{\ifmmode {\rm~cm}$^{-3}$ \else cm$^{-3}$\fi}
\newcommand\FWHM{\ifmmode {\rm~FWHM} \else ${\rm~FWHM}$\fi}
\newcommand\Msun{\ifmmode M_{\odot} \else $M_{\odot}$\fi}
\newcommand\Lsun{\ifmmode L_{\odot} \else $L_{\odot}$\fi}

\newcommand\hbeta{\ifmmode {\rm H}\beta \else H$\beta$\fi}
\newcommand\Kalpha{\ifmmode {\rm K}\alpha \else K$\alpha$\fi}
\newcommand\nh{\ifmmode N_{\rm H} \else N$_{\rm H}$\fi}

\title{{\it{XMM-Newton} view of a hard X-ray transient IGR~J17497--2821}}
\author[Alam et al.]{ \parbox[]{6.5in}{{
      Md. Shah Alam$^1\thanks{E-mail: alam@ctp-jamia.res.in}$,
      Dipanjan Mukherjee$^2$, Aditya S. Mondal$^3$, Gulab C. Dewangan$^4$,
      Sanjay Jhingan$^1$, Biplab Raychaudhuri$^2$}\\
      \footnotesize
    $^1$ Centre of Theoretical Physics, Jamia Millia Islamia,
New Delhi-110025, India\\
$^2$Research School of Astronomy and Astrophysics, The Australian National University, Canberra, ACT 2611, Australia \\
$^3$Department of physics, Visva-Bharati Santiniketan, West Bengal-731235, India\\
$^4$ IUCAA, Post Bag 4, Ganeshkhind, Pune 411007, India\\
}}

\date{\today}
\begin{document}
\pagerange{\pageref{firstpage}--\pageref{lastpage}} \pubyear{2013}

\maketitle
\label{firstpage}
\begin{abstract}
  We present spectral and energy dependent timing characteristics of
  the hard X-ray transient IGR~J17497--2821 based on {\it{XMM-Newton}}
  observations performed five and nine days after its outburst on 2006
  September 17. We find that the source spectra can be well described
  by a hard ($\Gamma\sim 1.50$) powerlaw and a weak multicolour disk
  blackbody with inner disk temperature $kT_{in}\sim 0.2\kev$. A broad
  iron K$\alpha$ line with FWHM$\sim27000{\rm~km~s^{-1}}$, consistent
  with that arising from an accretion disk truncated at large radius,
  was also detected. The power density spectra of IGR~J17497--2821,
  derived from the high resolution ($30\mu s$) timing mode \xmm{}
  observations, are characterised by broadband noise components that
  are well modelled by three Lorentzians. The shallow power law slope,
  low disk luminosity and the shape of the broadband power density
  spectrum indicate that the source was in the hard state. The rms
  variability in the softer energy bands ($0.3-2 \kev$) found to be
  $\sim 1.3$ times that in $2-5$ and $5-10 \kev$ energy bands. We also
  present the energy dependent timing analysis of the {\it RXTE}/PCA
  data, where we find that at higher energies, the rms variability
  increases with energy.
\end{abstract}


\section{Introduction}
The hard X-ray transient IGR~J17497--2821 \citep{2006ATel..885....1S}
was discovered by the IBIS telescope \citep{2003A&A...411L.131U}
on-board {\it INTEGRAL} mission \citep{2003A&A...411L...1W} on 2006
September 17 and subsequently observed by several other high energy
missions.  {\it Swift} observed the transient in two pointing on 2006
September 19 and 22, and refined the position to arcsec level
\citep{2007A&A...461L..17W}. The average {\it INTEGRAL} and {\it
  Swift}/XRT spectra were jointly well described by an absorbed hard
powerlaw ($\Gamma =1.67\pm0.06$) with a high energy cut-off at $\sim
200{\rm~keV}$.  {\it Suzaku} observed IGR~J17497--2821 eight days
after its discovery, during 2006 September 25--26, for a total of
about $53\ks$. \suzaku{} revealed the presence of an accretion disk
(kT$\sim 0.2 \kev$) \citep{Paizis}. Modelling the continuum with a
cutoff-powerlaw, the photon index was found to be $\sim 1.45$ with a
cutoff energy at $\sim 150 \kev$. A mild reflection component was also
detected in the spectrum \citep{Paizis}.  The Rossi X-ray timing
explorer {\it{RXTE}} observed this source during 2006 September 20--29
in seven pointings. From the {\it{RXTE}} follow ups, the source
spectrum was found to be constant with a powerlaw index
$\Gamma\sim1.55$. {\it{RXTE}} data are well represented by an absorbed
comptonized spectrum with an iron edge at $\sim 7\kev$
\citep{Rodriguez}. Timing analysis of the {\it RXTE} data revealed the
presence of three Lorentzian components in the power density spectra
of IGR~J17497--2821.  The source was observed by {\it{Chandra}} on
2006 October 1, for $19\ks$ during its decay phase when source flux
had dropped considerably. The position of IGR~J17497--2821 was
improved to $\alpha_{j2000}=17^{h}49^{m}38.037^{s}$,
$\delta_{j2000}=-28^{0}21^{'}17.37^{"}$. The {\it{Chandra}}
observation also found evidence for a cold accretion disk ($\sim 0.2
\kev$). There was a further hint of He-like Si-absorption line at
$1.867\kev$ \citep{Paizis07}.

Previously reported results indicate that the source never reached the
high/soft state during its outburst, and remained in the hard state
with a possible detection of an accretion disk component. However,
there are significant differences in the estimates of the powerlaw
photon index and the nature and extent of the accretion disk. Also,
detailed timing analysis of the available data has not been performed
before.

Among the existing high energy missions, X-ray Multi-Mirror Newton
(\xmm{}) is the only mission that provides good quality spectrum, and
high time resolution continuous light curves simultaneously in one of
the fast modes. In this paper, we utilize two \xmm{} observations of
IGR~J17497--2821 performed in the timing mode and investigate spectral
and temporal characteristics of the transient. In Sec.~\ref{sec.obs}
we discuss the details of the {\it XMM-Newton} observations and the
data reduction techniques used. We discuss the results of the analysis
of the {\it XMM-Newton} spectra in Sec.~\ref{sec.xmmspectra}. This is
followed in section~\ref{sec.timing} by the timing analysis of the
{\it XMM-Newton} data, and also the energy dependent timing analysis
of {\it RXTE} observations. In Sec.~\ref{sec.discuss} we summarise our
results from the spectral and timing analysis performed in this work.

\section{Xmm-Newton Observations and data reduction}\label{sec.obs}

\begin{figure}
  \centering
  \includegraphics[scale=0.30]{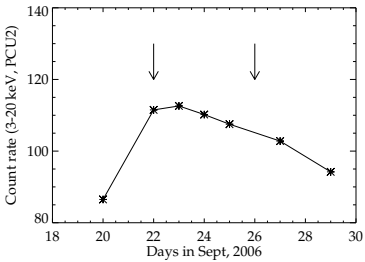}
  \caption{{\it RXTE/PCA} light curve of IGR~J17497-2821 from
    20th-29th September. Plotted are the mean net counts of PCU2 for
    each obsID. The time of the two {\it{XMM-Newton}} follow-up
    observations are marked by arrows.}\label{fig.pcu2}
\end{figure}

IGR~J17497--2821 was observed by {\it{XMM-Newton}} for the first time
on 2006 September 22 (obsID-0410580401, hereafter obs 1), for
approximately $33\ks$ (see Table~\ref{tab.obsdata}). A second
observation (obs.ID-0410580501, hereafter obs 2) was performed for 32
ks on 2006 September 26.  The two \xmm{} observations are marked on
the {\it RXTE/PCA} light curve in Fig.~\ref{fig.pcu2}. The
{\it{XMM-Newton}} Observatory \citep{Jansen} contains three $\rm 1500
cm^{2}$ X-ray telescopes, each with an European photon imaging camera
(EPIC) ($0.1-15 \kev$) at the focus. Two of the {\it{EPIC}} imaging
spectrometers use metal oxide semiconductor (MOS) CCDs~\citep{Turner}
and one uses pn CCDs~\citep{2001A&A...365L..18S}. Both the \xmm{}
observations of IGR~J17497--2821 provided the data from the three
EPICs MOS1, MOS2 and pn. The MOS1 camera was operated in the imaging
mode. The central CCD of MOS2 camera, containing the source, was
operated in the timing mode whereas other CCDs were operated in the
imaging mode. The EPIC-pn camera was operated in timing mode.

\begin{table*}
  \centering
  \begin{center}
    \caption{List of observations of IGR J17497-2821 by
      {\it{XMM-Newton}}}
    \begin{tabular}{| c | c | c | c | c |}
      \hline
      Obs.ID & Date of & Exposure  & Offset &  Count rate$^a$ \\
      &     observation                & Time (ks)      & (arcmin) & $ {\rm counts~s^{-1}}$ \\   
      \hline
      0410580401(obs 1) & 2006 Sep. 22 & 33 & 0.053 & $43.5$ \\
      0410580501(obs 2)& 2006 Sep. 26 & 32  & 0.051 & $39.4$\\
      \hline
    \end{tabular}\label{tab.obsdata}

\begin{tablenotes}
  \small
\item (a) Net count rate is taken from the {\it{XMM-Newton}} EPIC pn
  timing mode data after choosing a source region of width 15 pixel
  from the centre in the $0.3-10 \kev$ energy band and applying
  standard filtering criteria (see Sec.~\ref{sec.obs}).
\end{tablenotes}

\end{center}
\end{table*}

We processed the {\it{XMM-Newton}} observation data files, using the
science analysis software (SAS version 14.0), by applying the latest
calibration files available as on 9 January 2015. In the imaging mode
MOS1 data, the source was affected with severe pileup and we did not
use the MOS1 data further. The EPIC-pn camera and the central CCD of
MOS2 camera were operated in the timing mode in both the
observations. In this mode only one CCD chip is operated and the data
are collapsed into a 1-dimension row and read out at high speed, with
the second dimension replaced by timing information. This allows a
time resolution of $30\mu s$ for EPIC-pn and $1.75{\rm~ms}$ for
MOS2. From the EPIC-pn/MOS2 timing mode data we have created images of
the source using the SAS task `EVSELECT'. In the timing mode, the RAWY
co-ordinate gives the timing information and hence the source is
visible as a bright strip when plotting RAWX against RAWY.

To check whether the data were affected with soft proton flaring, the
light curve was extracted by selecting events with PATTERN$=$0 and
energy in the range $10-12\kev$. No evidence for flaring particle
background in the EPIC-pn and MOS2 data was found as the count rates
were steady. Due to this reason we did not apply any filtering. The
EPIC-pn cleaned event list was extracted by selecting events with
PATTERN$\le4$, FLAG$=0$ and energy in the $0.3$ to $10\kev$ range.
From the EPIC-pn cleaned event list, we extracted source events from a
$143.5{\rm~arcsec}$ (RAWX=20-55) wide box centred on the source
position. For background, we used a rectangular box with
RAWX=3-18. Using the SAS task EPATPLOT, we evaluated the pile-up
fraction for both the EPIC-pn data sets and found no significant
photon pile-up.  We used the source and background event lists to
create the source and background spectra for both the data sets. We
generated the response matrix and Ancillary response files using the
SAS task RMFGEN and ARFGEN, respectively.  We followed a similar
procedure to extract source and background spectra from the second
observation.  We rebinned the EPIC-pn spectra to over sample the FWHM
of the energy resolution by a factor of 5 and to have a minimum of 20
counts per bin.

For the MOS2 timing mode data, we used events with PATTERN=0, FLAG=0
and energy in the $0.3-10\kev$ range. We used rectangular regions of
width RAWX=270-338 ($74.8{\rm~arcsec}$) for the source and 260-270 for
the background and extracted the source and background spectra. We
have also extracted the background spectra from the MOS2 imaging mode
data.  The generation of the response files and grouping of the source
spectra were performed in the same way as done for EPIC-pn
data. Although it has been recommended to use the imaging mode
background for the MOS2 timing mode data
\footnote{\url{http://xmm2.esac.esa.int/docs/documents/CAL-TN-0018.pdf}
  (see XMM-SOC-CAL-TN-18, September 1, 2014), we have compared the
  MOS2 spectra by using both imaging mode background from the outer
  CCDs and timing-mode background from the central CCD.} We formed two
sets of MOS2 spectral data with the timing mode source spectrum in
combination with ($i$) the timing mode MOS2 background spectrum and
($ii$) imaging mode MOS2 background spectrum. We grouped both the data
sets in the same way as done for the EPIC-pn data, and compared the
two MOS2 data sets with different backgrounds. We fitted simple
absorbed powerlaw model to each of the two sets. We found that the two
MOS2 background spectra result in largely similar residuals. However,
in simultaneous EPIC-pn and MOS2 spectral fitting, we found slightly
larger discrepancy if we used the MOS2 spectral data with the imaging
mode background. Hence we have used the timing mode background
spectrum for further analysis. There still remains a small discrepancy
between the EPIC-pn and MOS2 spectra, which is consistent with
calibration uncertainty \citep{2014A&A...564A..75R}.

For the RGS data analysis, we used RGSPROC tool to reduce and extract
calibrated source and background spectrum and response files. We
extracted the lightcurve from the data contained in outer CCD (CCD
number=9) to check for particle background. GTI correction was applied
with expression rate$<=0.2$. Spectra and response files of 1st order
obtained from RGS1 and RGS2 were combined by using the SAS task
RGSCOMBINE. We grouped the spectra with minimum 15 counts per bin by
using GRPPHA.

\section{Spectral Analysis}\label{sec.xmmspectra}
We used XSPEC version 12.8.1g \citep{Arnaud} to perform the spectral
analysis. We begin with spectral analysis of the EPIC-pn and MOS2 data
obtained from the obs 1. We first fit the $2-10 \kev$ data with a
simple powerlaw model modified by the the {\scshape tbabs}
\citep{2000ApJ...542..914W} component to account for the galactic
absorption. Verner cross section
  \citep{1996ApJ...465..487V} and Wilm abundance
\citep{2000ApJ...542..914W} are applied in {\scshape
  tbabs}. A model {\scshape constant} is applied to account for
relative normalization between the instruments. Powerlaw index
$\Gamma$ is tied for the pn and MOS2 data. The simple powerlaw model
resulted in $\Gamma\sim1.5$ and $\chi^2=524.3$ for 416 degrees of
freedom (dof).  Extrapolating the powerlaw to low energies revealed a
broad excess below $2\kev$ which is likely the thermal emission from
an accretion disk. Performing the fit over the
$0.3-10\kev$ (pn and MOS2), the absorbed powerlaw
resulted in $\chi^2/dof=2801/558$. Addition of a
multicolour disk blackbody component {\scshape diskbb} \citep{Mit84}
improved the fit ($\chi^2/dof=1870/556$). The
  resulting fit showed a broad excess at ~1keV in the EPIC-pn data
  from both the observations.  However, the broad excess was not seen
  in the MOS2 data. To investigate this discrepancy, we attempted to
  fit the EPIC-pn, MOS2 and RGS data simultaneously. Due to lack of
  counts, we excluded RGS data below 1 keV.
 
\begin{figure*}
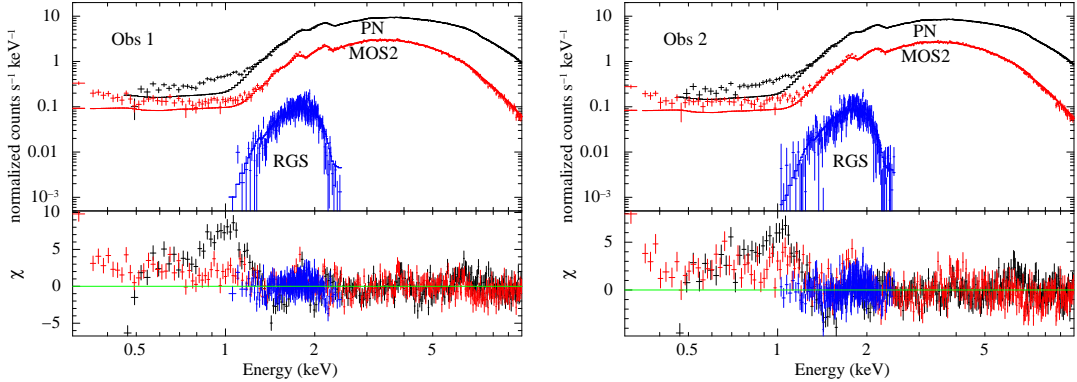

  \centering
  \includegraphics[scale=0.28,angle=-90]{obs1_cmb.ps}
  \includegraphics[scale=0.28,angle=-90]{obs2_cmb.ps}
  \caption{The EPIC-pn, MOS2 and RGS observed data, the model
    {\scshape const}$\times${\scshape tbabs$\times$(diskbb+powerlaw)}
    simultaneously fitted to these data sets. The residuals of obs 1
    (left panel) and obs 2 (right panel) show clear discrepancy
    between the these data sets below $\sim1.2\kev$.}
  \label{fig.pnmos}
\end{figure*}

Fig.~\ref{fig.pnmos} shows the observed EPIC-pn, MOS2 and RGS data,
folded model and the residuals. The datasets agree well
  above ~1.2 keV. Unfortunately, we cannot verify the 1-keV excess
  with the RGS data. However, similar excess in the timing mode data
have also been reported by several authors
\citep{Hiemstra,Boirin,Martocchia,Sala}.  We find that the EPIC-pn and
MOS2 data show excess emission below $\sim 1.2 \kev$ but with
different shapes.  The origin of this excess is not clearly understood
and may well be related to instrumental calibration.  We thus excluded
the $0.3-1.2\kev$ band data from the subsequent spectral analysis. We
have also excluded RGS data due to its poor signal to noise ratio.

The absorbed disk black-body plus powerlaw model fits the $1.2-10\kev$
data from obs 1, reasonably well. The reduced $\chi^2$ is $1.31$ for
$476$ degrees of freedom (d.o.f). Examination of the residuals from
this fit reveals an excess features like emission lines at
$\sim3.9\kev$ and $\sim 6.4\kev$. In order to model the structure
around $\sim 6.4\kev$, a Gaussian emission feature with an energy of
$6.3_{-0.1}^{+0.1} \kev$ and width ($\sigma$) of $<0.2 \kev$ was first
added to the existing model. This improves the fit with
$\Delta\chi^2=-37.2$ for three parameters. We also fit the data
obtained from the obs 2 with the same model, which resulted in
$\chi^2/dof=652.1/472$.  The feature $\sim 3.9\kev$ is very weak and
its exact nature is not clear to us.  The Gaussian line at $\sim
6.3\kev$ has substantial velocity width ($\sim 27000$\kms for obs 1
and obs 2) and can be readily identified with a broadened fluorescent
Iron K$\alpha$ emission line from the accretion disk.
To account for the relativistic broadening of the line, we replaced the Gaussian component at $\sim 6.3\kev$ with the {\scshape diskline} model \citep{Fabian}. This resulted in $\chi^2/dof=590.7/472$ for obs 1 (Table~\ref{tab.fitparams}, model 1) and $\chi^2/dof=651.2/471$ for obs 2. (Table~\ref{tab.fitparams} model 1) \\

To describe the continuum with a more physical model, we have also
fitted the continuum with the {\scshape{NTHCOMP}} model
\citep{1996MNRAS.283..193Z,1999MNRAS.309..561Z}. Fitting the
$1.2-10\kev$ data from obs 1 with the {\scshape
  tbabs$\times$(diskbb+nthcomp)} model provided an acceptable fit with
$\chi^2/dof=630.8/476$. Here the fitting residuals also showed the
same emission features around the same energy. In order to investigate
the relativistic broadening of the line, we replaced the Gaussian line
with the {\scshape diskline} model. This resulted in
$\chi^2/dof=592.2/472$ for obs 1 (Table~\ref{tab2.fitparams}, model 4)
and $\chi^2/dof$=651.6/471 for obs 2 (Table~\ref{tab.fitparams}, model
4). We fixed the emissivity index ($\beta$) and the outer radius of
the disk to the value $-3.0$ and $1000r_{g}$ respectively for both the
observations. The best fit parameter values for different models are
reported in Table~\ref{tab2.fitparams} and \ref{tab.fitparams} for obs
1 and 2 respectively.

We have also checked for the variation of iron line energy and disc
black body parameters by using different continuum models such as
POWERLAW, SIMPL \citep{2009PASP..121.1279S}, {\scshape COMPTT}
\citep{1995ApJ...449..188H,1995ApJ...450..876T} and {\scshape
  NTHCOMP}. We find that the best-fit parameters of the iron line and
the thermal accretion disc are consistent within error for the
different continuum models (Table~\ref{tab2.fitparams} and
\ref{tab.fitparams} ).

\begin{figure*}
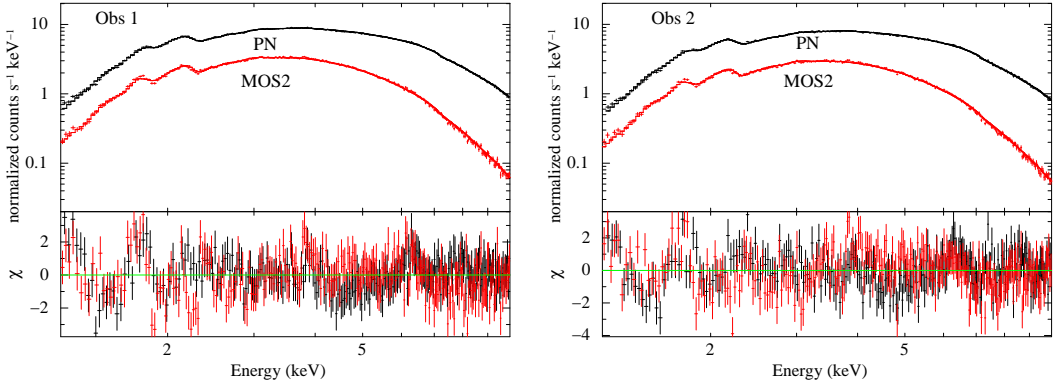

  \centering

  \includegraphics[scale=0.28,angle=-90]{obs1_res.ps} 
  \includegraphics[scale=0.28,angle=-90]{obs2_res.ps}
  \caption{The EPIC-pn and MOS2 observed data and the best-fit
    {\scshape const}$\times${\scshape tbabs$\times$(diskbb+powerlaw)}
    model (top panel) and residuals (bottom panel) for obs 1 (left)
    and obs 2 (right). The residuals show possible presence of
    emission features at $\sim 6.4 \kev$.}
  \label{fig.pncont1}
\end{figure*}

\begin{figure*}
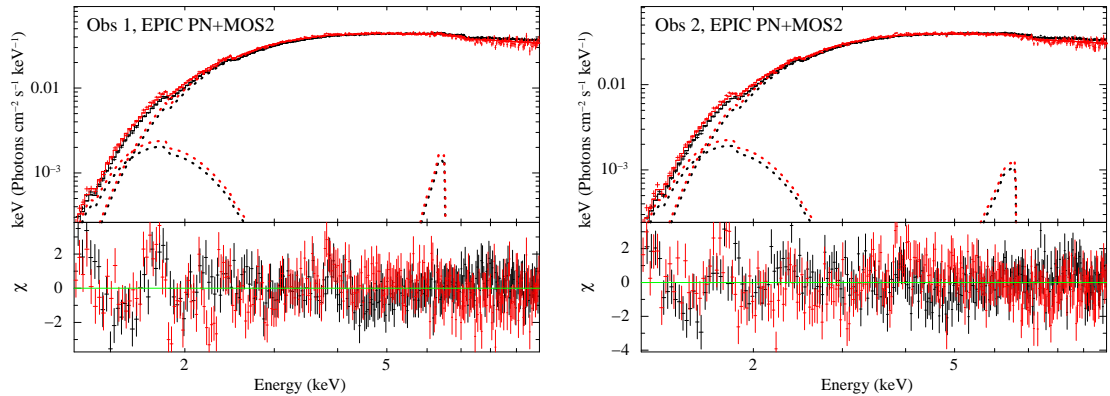

  \centering

  \includegraphics[scale=0.29,angle=-90]{obs1_diskl.ps} 
  \includegraphics[scale=0.29,angle=-90]{obs2_diskl.ps}
  \caption{The unfolded EPIC-pn+MOS2 spectral data and the
    best-fitting model {\scshape
      tbabs$\times$(diskbb+nthcomp+diskline)} (see
    Table~(\ref{tab2.fitparams} and \ref{tab.fitparams}) for best-fit
    parameters) and deviations of the observed data from the
    model. Emission line at $\sim 6.6 \kev$ (obs 1 and 2) can be
    identified as the broad iron K$\alpha$ line modelled as {\scshape
      diskline}.}
  \label{fig.unfolded}
\end{figure*}

\section{Temporal Analysis}\label{sec.timing}
\subsection{The timing mode {\it\textbf{XMM-Newton}} EPIC-pn
  data}\label{sec.timing.xmm}
The timing mode EPIC-pn data with a time resolution of $30\mu s$ are
well suited to study temporal behaviour of IGR~J17497--2821. We used
the General High-energy Aperiodic Timing Software (GHATS) software
package (version
1.1.0)\footnote{\url{http://astrosat.iucaa.in/~astrosat/GHATS_Package/Home.html}}
for our timing analysis.  We first generate the power density spectra
(PDSs) over the entire $0.3-10 \kev$ band from the source event lists
for both the observations.  The data were divided into time segments
of length $786\s$, and a Fourier transform was performed with the
Nyquist frequency set at $\sim 42\hz$. The resulting PDSs were Poisson
noise corrected \citep{Klis1995a} and logarithmically rebinned. The
PDSs were evaluated using rms normalisation
\citep{Hasinger1990b,Klis1995b}.

\begin{figure*}
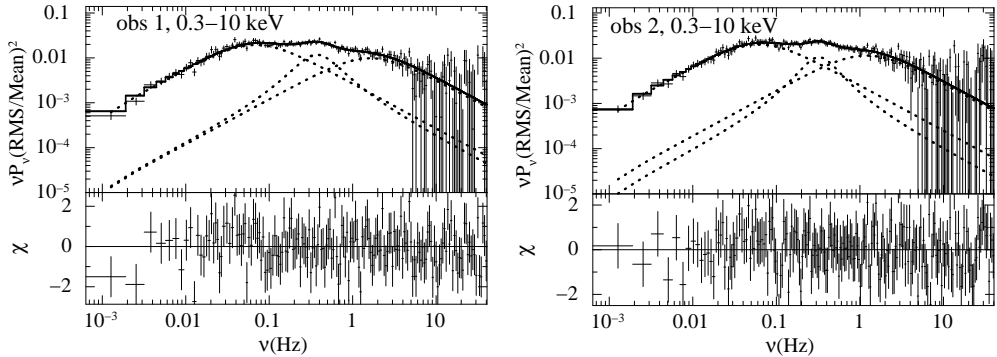

    \centering

    \includegraphics[scale=0.25,angle=-90]{obs1_0.3_10.ps}
    \includegraphics[scale=0.25,angle=-90]{obs2_0.3_10.ps}
    \caption{The \xmm{} power density spectra of obs 1 and obs 2 of
      IGR~J17497--2821 derived from the full band EPIC-pn fitted with
      three lorentzians (top panel) and residuals (bottom panel). The
      PDSs were derived from the high time resolution ($30\mu\s$)
      EPIC-pn data obtained from obs 1 and obs 2 with
      \xmm{}.}\label{fig.xmm.timing}
  \end{figure*}

Following \citep{Belloni02,Nowak}, we fit the broadband noise
components in the PDSs using Lorentzian models with the following
expression
\begin{equation}
  P(\nu) = \frac{r^2 \Delta}{\pi} \frac{1}{\Delta^2 + (\nu-\nu_0)^2}
\end{equation}
where P($\nu$) is the power spectral density, $r$ the integrated
fractional rms, $\Delta$ is the half width at half of the maximum
power (HWHM) and $\nu_{0}$ the centroid frequency.  As is typical of
blackhole binaries \citep{Belloni02,Nowak} we attempt to fit the PDS
with three Lorentzian components (as shown in
Fig.~\ref{fig.xmm.timing}): a broad zero-centred low frequency noise
component (LFN), a middle component (mid-frequency noise or MFN) and a
broad zero-centred component to account for the noise beyond $1\hz$
(high frequency noise or HFN). A similar approach was followed in a
previous work on the {\it RXTE} timing analysis by \citet{Rodriguez}
(JR07). The fit results are presented in Table~\ref{table.xmm}. The
total rms variability was found to be $\sim 32\%$ (frequency range
$0.01-40$ Hz).

The shape and nature of the power spectra and rms variability are
often found to vary with energy \citep{lin2000,Belloni1997}. Such
trends gives important clues to the nature of the intrinsic emission
processes and cause of the observed variability. Hence we have divided
$0.3-10 \kev$ band EPIC-pn data into three different energy bands
$0.3-2$, $2-5$ and $5-10\kev$, for the timing analysis. The main
motivation of performing the energy dependent timing analysis was to
identify if the timing characteristics are correlated with the
spectral parameters. The choice of the energy bands are indeed
empirical to some extent. The low energy band (0.3-2 keV) was selected
to sample the photons which might pre-dominantly originate from the
accretion disk.
The highest count rate was found to be in the $2-5\kev$ band, whereas
the $0.3-2\kev$ band had the lowest count rate (see
table~\ref{table.xmm}). Signal to noise ratio of the $2-5\kev$ band
data was found to be approximately three times more than that of
$0.3-2\kev$ band for both observations.

We have computed the PDSs of three the energy bands ($0.3-2$ , $2-5$
and $5-10 \kev$) following the method as described earlier. We
consider only those components which are detected with $3\sigma$
significance or above\footnote{Obtained by dividing the norm of the
  component with the lower limit of the error computed with 1$\sigma$
  confidence interval}. We find that the MFN component is detected
significantly only at high energy bands ($2-5$ keV and $5-10$ keV).

\begin{figure*}
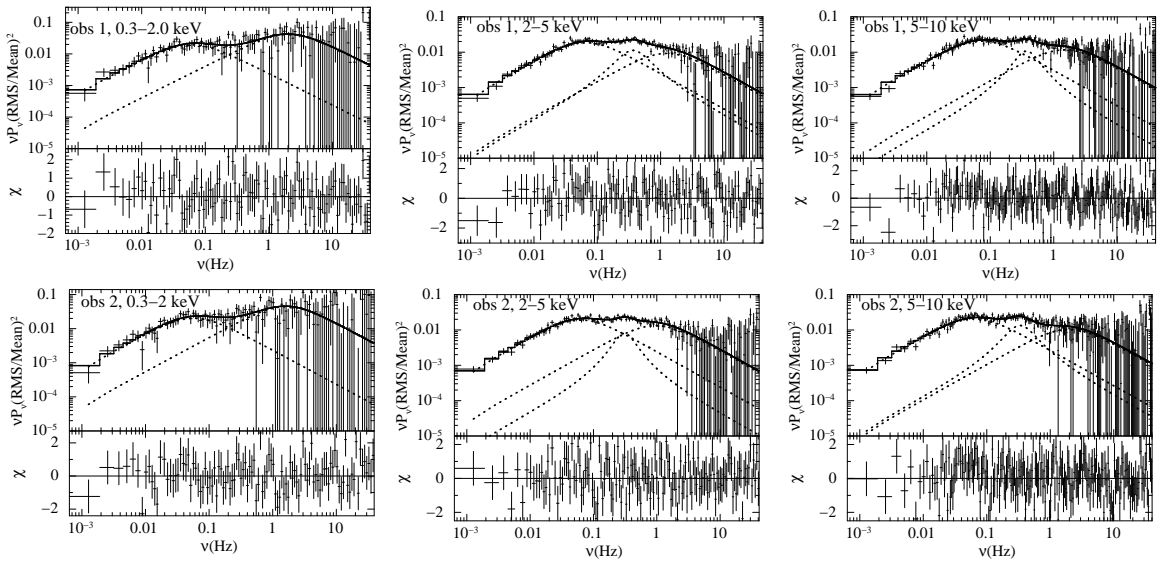

    \centering \vspace{0cm}

    \includegraphics[scale=0.19,angle=-90]{obs1_0.3_2.ps} 
    \includegraphics[scale=0.19,angle=-90]{obs1_2_5.ps} 
    \includegraphics[scale=0.19,angle=-90]{obs1_5_10.ps}\\

    \includegraphics[scale=0.19,angle=-90]{obs2_0.3_2.ps}
    \includegraphics[scale=0.19,angle=-90]{obs2_2_5.ps}
    \includegraphics[scale=0.19,angle=-90]{obs2_5_10.ps}\\
    \caption{ The power density spectra of IGR~J17497--2821 in
      different energy bands fitted with lorentzian(top panel of
      plots) and residual (bottom panel of plots). The PDSs were
      derived from the high time resolution ($30\mu\s$) EPIC-pn data
      obtained from obs 1 and obs 2 with \xmm{}.  Sum of two
      Lorentzians are applied to the PDSs in the $0.3-2\kev$ band and
      sum of 3 Lorentzians are applied for the PDSs in the $2-5$ and
      $5-10\kev$ bands.}\label{fig.xmm.multie}
  \end{figure*}

For obs 1, the $0.3-2\kev$ band PDS fit well with only the LFN and HFN
components ($\chi ^2 /{\rm dof} \sim 88/104$).  The PDS of the $2-5
\kev$ energy band was first fit with two zero centred Lorentzian (LFN
and HFN) as before, with $\chi^{2}$/dof=158.31/136. Introducing of a
middle component at $\sim$ 0.3 Hz improves the fit
($\chi^{2}/dof=140.95/133$), with the significance of the resulting
MFN being $\sim 3 \sigma$.  The PDS of the $5-10 \kev$ band similarly
required all three Lorentzian components for a satisfactory fit (as
shown in Fig.~\ref{fig.xmm.multie}).

To test the possible presence of an MFN component in the $0.3-2$ keV
band, we applied a Lorentzian with centroid and width fixed to that of
the MFN component of the $2-5$ keV band. We varied the norm of the
applied Lorentzian to obtain a fit to the PDS. The resulting
improvement in $\chi ^2$ was marginal with $\Delta \chi ^2 = -3$ for
one additional parameter.  with the component being $1.5 \sigma$
significant. Computing the errors with 90\% confidence limit, we find
the strength of the norm to be $\sim 0.014^{+0.014}_{-0.014}$. Thus
the lower limit is consistent with zero confirming that the component
may not be significant.  However, the upper limit is comparable to the
strength of the MFN in the $2-5$ keV band (see
Table~\ref{table.xmm}). Quality of the $0.3-2$ keV data is poor as
compared to the data in the $2-5$ keV. This also affects the
significance of MFN in low energy band $0.3-2$ keV. This indicates
that although the MFN component may not be significant, but we can not
rule out its presence given the current SNR of the data.

The PDSs of obs 2 were fit in a similar fashion as described above
(see Table~\ref{table.xmm}). For both observations, integrated rms
values (frequency range 0.01-40 Hz) for PDSs of energy bands $0.3-2$,
$2-5$, $5-10$ and $0.3-10 \kev$ are reported in Table~\ref{table.xmm}.

\subsection{{\it\textbf{RXTE}}/PCA data}
To test the energy dependence of the PDS over the course of the
outburst we have analysed the available {\it RXTE} data in different
energy channels. {\it{RXTE}} monitored this source from 20-29
September, 2006. For south Atlantic anomaly (SAA) correction, we
removed 30 minutes of data. In this duration the satellite was exposed
to strong radiation. Number of PCU which observe the source is
different for different observations. 92016-01-01-00, 92016-01-01-02,
92016-01-02-01 and 92016-01-02-02 have observed the source with 4, 1,
1 and 2 PCU respectively.  In this work we have performed the timing
analysis for all the four observations with GOODXENON event files.
The {\it RXTE/PCA} timing analysis was performed for the energy bands
$2-5$, $5-10$ and $10-20 \kev$, respectively with GHATS V 1.1.0. The
PDSs were created for time segments of length 128s with a Nyquist
frequency of 64 Hz and deadtime corrections were applied
\citep{1995ApJ...449..930Z}. Logarithmic rebinning was applied to the
Poisson noise corrected PDS.  The resulting PDSs were fit with
Lorentzian components as described in Sec.~\ref{sec.timing.xmm}. We
present in Fig.~\ref{fig.xte.pds} the PDSs of different energy bands
of the {\it RXTE} observations and describe the results below.

We found that a significant MFN component was detected only in the
$5-10$ kev and $10-20$ keV of obsID 92016-01-01-00 (20th
September). The $2-5 \kev$ band was first fit with the LFN (low
frequency noise) and HFN (high frequency noise) components
($\chi^{2}$/dof=141.77/136). The remaining residual structures at
$\sim 0.3$ Hz was then fit with the MFN (mid-frequency noise)
component to get the final $\chi^{2}$/dof to be 128.86/133. The middle
broad noise component was found to be only $\sim$ 2.67$\sigma$
significant, and hence discarded from the final results as we retain
only those components with a detection significance above $\sim
3\sigma$. Signal to noise ratio of the data in the $2-5$ keV is found
to be 1.3 times smaller than the $5-10$ keV band. So the presence of
MFN can not be discarded as lower significance of MFN may be due to
bad quality data of $2-5$ keV band. The PDSs of the $5-10$ and $10-20
\kev$ energy ranges were fit in a similar fashion with three
Lorentzian components. In both cases the middle component was required
for the fit with a significance of $\sim 4.0\sigma$ and $\sim
5.4\sigma$, respectively (see Table~\ref{pds.rxte1}). Thus similar to
the results of \xmm{} data, a statistically significant MFN component
was detected only at the higher energy bands.  The PDS of other obsIDs
analysed fit well with only two lorentzian components, with the MFN
being not statistically significant.

\begin{table*}
  \caption{Best-fit spectral model parameters of IGR~J17497--2821 based on the combined spectral analysis of EPIC-pn and MOS2 data sets from obs1.} 
  \begin{tabular}{|p{1.7cm}|p{2.5cm}|p{2.5cm}|p{2.4cm}|p{2.6cm}|p{2.4cm}|}
    \hline
    &  & \multicolumn{3}{c}{obs1 (obs.ID : 0410580401)} & \\
    \hline
     
    Model component & Parameter & model 1 & model 2 & model 3 & model 4 \\
    \hline
    {\scshape constant}&&1.0(fixed)(pn)&1.0(fixed)(pn)&1.0(fixed)(pn)&1.0(fixed)(pn)\\
    && ${0.9_{-0.004}^{+0.004}}$(MOS2)&${0.7_{-0.008}^{+0.007}}$(MOS2)&${0.9_{-0.004}^{+0.004}}$(MOS2)&${0.9_{-0.004}^{+0.004}}$(MOS2)\\
    {\scshape tbabs} & $ N_{H}$($\times 10^{22}{\rm~cm^{-2}}$) & ${7.5_{-0.1}^{+0.1}}$ & ${7.5_{-0.1}^{+0.1}}$ &$7.5_{-0.1}^{+0.1}$ &$7.6_{-0.1}^{+0.1}$  \\
    {\scshape diskbb} & kT$_{in}$(keV) & $0.2_{-0.006}^{+0.006}$ & $0.2_{-0.006}^{+0.006}$ &$0.2_{-0.006}^{+0.006}$ &$0.2_{-0.006}^{+0.006}$  \\
    & Norm{($\times10^{5}$)}   & $2.9_{-0.7}^{+1.0}$ &$3.1_{-0.8}^{+1.0}$  & $2.8_{-0.7}^{+1.0}$ &$3.3_{-0.8}^{+1.1}$  \\
    {\scshape diskline}   & $E_{line}$ (keV)& $6.5_{-0.3}^{+0.2}$ &$6.6_{-0.4}^{+0.1}$ & $6.5_{-0.3}^{+0.2}$ &$6.6_{-0.4}^{+0.2}$ \\
    &  $\beta$ &$-3.0(fixed)$ &$-3.0(fixed)$ &$-3.0(fixed)$ &$-3.0(fixed)$ \\
    & $r_{in}$($r_g$) &$14.6_{-6.5}^{+952.1}$ &$11.5_{-3.8}^{+203.6}$ &$20.3_{-11.4}^{+968.6}$ &$14.3_{-6.5}^{+235.3}$ \\
    & $r_{out}$($r_g$) &$1000(fixed)$ &$1000(fixed)$  &$1000(fixed)$ & $1000(fixed)$ \\
    & Inclination ($\theta$) & $13.8_{peg}^{+25.6}$ & $15.2_{peg}^{+43.0}$ & $11.2_{peg}^{+35.2}$ & $13.4_{peg}^{+41.8}$ \\
    & $f_{line}{(\times10^{-4})}$ & $1.5_{-0.6}^{+0.7}$ &$1.9_{-0.8}^{+0.7}$ & $1.2_{-0.4}^{+0.9}$ &$1.6_{-0.6}^{+0.8}$ \\
    {\scshape powerlaw}  & $\Gamma$ &$1.5_{-0.01}^{+0.01}$& && \\
            
    & Norm &  $0.1_{-0.002}^{+0.002}$ &  &&  \\
    {\scshape simpl} & $\Gamma$ & &$1.5_{-0.02}^{+0.01}$ & & \\
    & Fracsctr &  & $0.03_{-0.005}^{+0.004}$ & & \\
    {\scshape comptt } & T0(keV)&&& $0.2_{-0.006}^{+0.006}$~$^a$ & \\
    & $\tau$ && &$0.7_{-0.02}^{+0.02}$ &  \\
    &norm($\times 10^{-3}$)&&&$3.8_{-0.05}^{+0.06}$&\\
      
    {\scshape nthcomp } & $\Gamma$ &&&&$1.50_{-0.01}^{+0.01}$ \\
    & kT$_{bb}$(keV) && & &$0.2_{-0.006}^{+0.006}$~$^b$ \\
    & Norm &&&&$0.1_{-0.002}^{+0.002}$ \\

    \hline 
    & $\chi^{2}/dof$ &$590.7/472$ &$651.2/472$ &$590.0/472$ &$592.2/472$  \\
    \hline
  \end{tabular}\label{tab2.fitparams} \\
  We used {\scshape const$\times$tbabs$\times$(continuum+diskbb+diskline)} to fit spectrum.  For CONTINUUM component, we applied {\scshape powerlaw, simpl, comptt} and {\scshape nthcomp} in model 1, 2, 3 and 4 respectively.\\
  (a)and (b): parameters tied with kT$_{in}$ parameter of DISKBB.\\
\end{table*}
\begin{table*}
  \caption{Best-fit spectral model parameters of IGR~J17497--2821 based on the combined spectral analysis of EPIC-pn and MOS2 data sets from obs2.} 

\begin{tabular}{|p{1.7cm}|p{2.5cm}|p{2.5cm}|p{2.4cm}|p{2.6cm}|p{2.4cm}|}
  \hline
  &  & \multicolumn{3}{c}{obs1 (obs.ID : 0410580501)} & \\
  \hline
     
  Model component & Parameter & model 1 & model 2 & model 3 & model 4 \\
  \hline
  {\scshape constant}&&1.0(f)(pn)&1.0(f)(pn)&1.0(f)(pn)&1.0(f)(pn)\\
  && ${0.9_{-0.004}^{+0.004}}$(MOS2)&${0.7_{-0.008}^{+0.006}}$(MOS2)&${0.9_{-0.004}^{+0.004}}$(MOS2)&${0.9_{-0.004}^{+0.004}}$(MOS2)\\
  {\scshape tbabs} & $ N_{H}$($\times 10^{22}{\rm~cm^{-2}}$) & ${7.6_{-0.1}^{+0.1}}$ & ${7.5_{-0.1}^{+0.1}}$ &$7.5_{-0.1}^{+0.1}$ &$7.7_{-0.1}^{+0.05}$  \\
  {\scshape diskbb} & kT$_{in}$(keV) & $0.2_{-0.006}^{+0.005}$ & $0.2_{-0.004}^{+0.004}$ &$0.2_{-0.006}^{+0.006}$ &$0.2_{-0.006}^{+0.005}$  \\
  & Norm($\times10^{5}$)   & $3.5_{-0.8}^{+1.1}$ &$3.5_{-0.8}^{+1.3}$  & $3.4_{-0.8}^{+1.1}$ &$4.0_{-0.9}^{+1.3}$  \\
  {\scshape diskline}   & $E_{line}$ (keV)& $6.6_{-0.5}^{+0.2}$ &$6.6_{-0.2}^{+0.1}$ & $6.6_{-0.5}^{+0.2}$ &$6.6_{-0.2}^{+0.2}$ \\
  &  $\beta$ &$-3.0(fixed)$ &$-3.0(fixed)$ &$-3.0(fixed)$ &$-3.0(fixed)$ \\
  & $r_{in}$($r_g$) &$11.5_{-3.8}^{+80.1}$ &$10.8_{-4.1}^{+40.3}$ &$11.7_{-3.8}^{+90.0}$ &$11.2_{-3.8}^{+41.8}$ \\
  & $r_{out}$($r_g$) &$1000(fixed)$ &$1000(fixed)$  &$1000(fixed)$ & $1000(fixed)$ \\
  & Inclination ($\theta$) & $13.2_{peg}^{+31.7}$ & $15.1_{peg}^{+12.9}$ & $13.7_{peg}^{+46.8}$ & $13.9_{peg}^{+11.5}$ \\
  & $f_{line}{(\times10^{-4})}$   & $1.5_{-0.5}^{+0.6}$ &$1.9_{-0.5}^{+0.6}$ & $1.5_{-0.5}^{+0.6}$ &$1.6_{-0.5}^{+0.7}$ \\
  {\scshape powerlaw}  & $\Gamma$ &$1.5_{-0.01}^{+0.01}$& && \\
            
  & Norm &  $0.1_{-0.002}^{+0.002}$ &  &&  \\
  {\scshape simpl} & $\Gamma$ & &$1.5_{-0.01}^{+0.01}$ & & \\
  & Fracsctr &  & $0.02_{-0.004}^{+0.004}$ & & \\
  {\scshape comptt } & T0(keV)&&& $0.2_{-0.006}^{+0.006}$~$^a$ & \\
  & $\tau$ && &$0.8_{-0.02}^{+0.02}$ &  \\
  &Norm($\times 10^{-3}$)&&&$3.5_{-0.05}^{+0.05}$&\\
      
  {\scshape nthcomp } & $\Gamma$ &&&&$1.5_{-0.01}^{+0.01}$ \\
  & kT$_{bb}$(keV) && & &$0.2_{-0.006}^{+0.005}$~$^b$ \\
  & Norm &&&&$0.1_{-0.001}^{+0.002}$ \\

  \hline 
  & $\chi^{2}/dof$ &$651.2/471$ &$741.7/471$ &$651.3/471$ &$651.5/471$  \\
  \hline
\end{tabular}\label{tab.fitparams} \\
We used {\scshape const$\times$tbabs$\times$(continuum+diskbb+diskline)} to fit spectrum.  For CONTINUUM component, we applied powerlaw, simpl, comptt and nthcomp in model 1, 2, 3 and 4 respectively.\\
(a)and (b): parameters tied with kT$_{in}$ parameter of DISKBB.\\
\end{table*}

\begin{table*}
  \caption{Results of fits to the PDSs of IGR~J17497--2821 derived from the timing mode EPIC-pn data. Frequency range 0.01-40 Hz has been used to calculate the integrated rms.}  \label{table.xmm}
  \begin{tabular}{|c|c|c|c|c|c|c|}
    \hline 
    Model component~$^a$ &Parameter~$^b$ & $0.3-2 \kev$ & $2-5 \kev$ & $5-10\kev$ & $0.3-10 \kev$  \\
    \hline
    &  & & \multicolumn{2}{c}{obs 1 (obs.ID : 0410580401)} &&\\
 
    LFN &2$\Delta$ (Hz)  & $0.1_{-0.02}^{+0.03}$ & $0.1 \pm 0.01$ & $0.1_{-0.01}^{+0.01}$&$0.1 \pm 0.01$  \\

    & Norm              &$6.07^{+0.9}_{-0.9} \times 10^{-02}$ &$5.8^{+0.4}_{-0.5} \times 10^{-02}$  & $6.4^{+0.3}_{-0.4} \times 10^{-02}$  &$6.1^{+0.3}_{-0.4} \times 10^{-02}$ \\  
 
 
    HFN &2$\Delta$ (Hz)  & $4.1_{-1.4}^{+2.0}$ & $2.5_{-0.9}^{+1.7}$ & $3.0_{-0.8}^{+1.0}$& $3.3_{-0.6}^{+0.7}$ \\

    & Norm              & $0.1^{+0.03}_{-0.03}$        &$2.9^{+0.9}_{-1.0} \times 10^{-02}$ &$3.7^{+0.5}_{-0.6} \times 10^{-02}$ &$3.1^{+0.3}_{-0.4} \times 10^{-02}$ \\

    MFN & $\nu_{0}$ (Hz)& -- & $0.3_{-0.08}^{+0.05}$ &$0.3_{-0.05}^{+0.04}$ &$0.3_{-0.05}^{+0.04}$ \\
 
    &2$\Delta$ (Hz)  & --  & $0.4_{-0.2}^{+0.2}$ & $0.3_{-0.1}^{+0.2}$ & $0.4_{-0.1}^{+0.1}$\\
 
    &Norm           & -- & $1.8^{+1.3}_{-1.0} \times 10^{-02}$&$1.4^{+0.7}_{-0.5} \times 10^{-02}$& $2.0^{+0.7}_{-0.5} \times 10^{-02}$ \\  

    &Significance ($\sigma$)$^c$  & -- & 2.8 & 3.9 & 5.9\\
    &Integrated rms (in percentage)&$43.0^{+1.0}_{-1.0}$&$32.0^{+0.3}_{-0.3}$&$33.0^{+0.4}_{-0.4}$&$32.3^{+0.2}_{-0.2}$\\ 
    &mean count rate(counts/sec)$^d$  & 2.8 & 23.3 & 17.8 & 43.8\\
    &$\chi^2/dof$ & 88.0/104 & 140.9/133 & 200.1/192 & 152.0/163  \\
    \hline
    &  & & \multicolumn{2}{c}{obs 2 (obs.ID : 0410580501)} &&\\
 
    LFN &2$\Delta$ (Hz)  & $0.1_{-0.02}^{+0.03}$ &$0.1 \pm 0.01$ & $0.1_{-0.01}^{+0.01}$& $0.1 \pm 0.01$ \\

    & Norm              &$6.3^{+1.0}_{-1.0} \times 10^{-02}$ & $6.0^{+0.3}_{-0.3} \times 10^{-02}$ &$6.5^{+0.4}_{-0.4} \times 10^{-02}$ &$6.3^{+0.3}_{-0.3} \times 10^{-02}$ \\

    HFN &2$\Delta$ (Hz)  & $3.2_{-1.2}^{+2.1}$ & $1.9_{-0.4}^{+0.6}$ & $3.9_{-1.2}^{+1.8}$ & $2.4_{-0.4}^{+0.5}$\\

    & Norm  & $0.1^{+0.04}_{-0.03}$&$4.1^{+0.5}_{-0.7} \times 10^{-02}$  &  $3.3^{+0.5}_{-0.6} \times 10^{-02}$ &$3.6^{+0.4}_{-0.4} \times 10^{-02}$ \\
 
    MFN& $\nu_{0}$ (Hz)& --  & $0.3_{-0.04}^{+0.03}$ &$0.3_{-0.04}^{+0.03}$ &$0.3_{-0.04}^{+0.03}$ \\
 
    &2$\Delta$ (Hz)  & --  & $0.2_{-0.1}^{+0.2}$ & $0.4_{-0.1}^{+0.1}$ & $0.3_{-0.09}^{+0.1}$\\

    &Norm           & -- &$9.3^{+8.0}_{-4.9} \times 10^{-03}$ & $2.0^{+0.8}_{-0.6} \times 10^{-02}$ & $1.5^{+0.6}_{-0.5} \times 10^{-02}$ \\  
  
    &Significance ($\sigma$)$^c$  & -- & 2.9 & 5.0 & 5.0\\
    &Integrated rms (in percentage)&$44.0^{+1.8}_{-1.5}$&$33.0^{+0.3}_{-0.2}$&$34.0^{+0.5}_{-0.4}$&$32.6^{+0.2}_{-0.2}$\\ 
    &mean count rate(counts/sec)$^d$  & 2.6 & 21.2 & 16.2 & 39.9\\
    &$\chi^2/dof$ & 90.9/104 & 130.8/133 & 207.1/192 & 181.2/192  \\
    \hline
  \end{tabular}

  (a) LFN and HFN are Lorentzian components used to model the  low and high frequency noise. The centroid frequencies of the LFN and HFN components were frozen to zero. MFN is the mid frequency Lorentzian used to fit broad feature between low and high frequency noise components in the PDS. Sum of 2 Lorentzians are applied for energy band 0.3-2keV, whereas sum of 3 Lorentzians are applied for energy bands 2-5, 5-10 and 0.3-10keV. \\
  (b) $\nu_{0}$ and 2$\Delta$ are the centroid frequency and FWHM of the Lorentzian respectively. \\
  (c) The significance of the middle component computed with 1 $\sigma$ error is presented separately. \\
  (d) mean count rate is calculated from lightcurves in the $0.3-2$,
  $2-5$, $5-10$ and $0.3-10\kev$ band after applying standard filtering and taking source region of 15 pixel width.
\end{table*}

  \begin{table*}
    \caption{Results of fits to the PDS of {\it RXTE/PCA}  observations on 2006 20th and 23rd September. Frequency range 0.01-64Hz is used to calculate the rms.}\label{pds.rxte1}
    \begin{center}
      \begin{tabular}{|c|c|c|c|c|c|c|}
        \hline
        obsID & Model component & Parameter & $2-5\kev$ & $5-10\kev$ & $10-20 \kev$   \\
        \hline

        &LFN & 2$\Delta$ (Hz)  & $0.1_{-0.02}^{+0.02}$ & $0.1_{-0.02}^{+0.02}$ & $0.1_{-0.02}^{+0.02}$ \\
        &    &  Norm          & $2.6^{+0.4}_{-0.4} \times 10^{-02}$ & $2.7^{+0.2}_{-0.3}\times 10^{-2}$ & $4.2^{+0.4}_{-0.4}\times 10^{-2}$ \\

	92016-01-01-00 & HFN &2$\Delta$ (Hz)  &$1.0_{-0.2}^{+0.3}$ & $2.9_{-0.7}^{+1.1}$ & $3.7_{-0.9}^{+1.2}$ \\
        (20 sep 2006)  &     & Norm & $2.3 \pm 0.3 \times 10^{-2}$ & $1.5^{+0.3}_{-0.3}\times 10^{-2}$ & $2.0^{+0.3}_{-0.3}\times 10^{-2}$ \\

        &MFN & $\nu_{0}$ (Hz)&  &$0.3_{-0.08}^{+0.05}$ &$0.3_{-0.05}^{+0.04}$ \\
        &&2$\Delta$ (Hz)  & & $0.4_{-0.1}^{+0.2}$ & $0.3_{-0.09}^{+0.1}$ \\
        &    &  Norm          &  & $1.0^{+0.6}_{-0.4}\times 10^{-2}$ & $1.4^{+0.5}_{-0.4}\times 10^{-2}$ \\
        &&Significance ($\sigma$)$^a$  & & 4.0 & 5.4 \\
        &&Integrated rms (in percentage)&$21.3^{+0.2}_{-0.3}$&$22.3^{+0.2}_{-0.3}$&$26.7^{+0.3}_{-0.3}$\\

        && $\chi^2/dof$ & 141.8/136 & 159.6/156 & 220.1/201   \\
            
        \hline

        &LFN & 2$\Delta$ (Hz)  & $7.5^{+2.7}_{-2.0} \times 10^{-2}$ &$7.8^{+2.8}_{-2.1} \times 10^{-2}$&$8.3^{+2.8}_{-2.1} \times 10^{-2}$ \\
        &    &  Norm          & $2.0 \pm 0.5 \times 10^{-2}$ &$2.3 \pm 0.5 \times 10^{-2}$ & $3.3^{+0.7}_{-0.7}\times 10^{-2}$ \\

 	92016-01-01-02 & HFN &2$\Delta$ (Hz)  & $0.9_{-0.2}^{+0.3}$ & $0.9_{-0.2}^{+0.2}$ & $1.0_{-0.2}^{+0.3}$ \\
        (23 sep 2006)  &     & Norm & $2.9^{+0.5}_{-0.6} \times 10^{-2}$ & $3.5^{+0.4}_{-0.4}\times 10^{-2}$ & $4.4^{+0.6}_{-0.6}\times 10^{-2}$ \\
        &&Integrated rms (in percentage)&$21.1^{+0.7}_{-0.7}$&$23.1^{+0.4}_{-0.4}$&$26.6^{+0.6}_{-0.7}$\\

        && $\chi^2/dof$ & 129.0/112 & 114.6/112 & 208.7/204   \\
        \hline

      \end{tabular}
    \end{center}
    {\small (a) The significance of the middle component computed with 1 $\sigma$ error is presented separately.} \\
  \end{table*}

  \begin{table*}
    \caption{Results of fits to the 0.01-64 Hz PDS of {\it RXTE/PCA}  observations on 24th and 25th September, 2006.}\label{pds.rxte2}
    \begin{center}
      \begin{tabular}{|c|c|c|c|c|c|c|}
        \hline
        obsID & Model component & Parameter & $2-5\kev$ & $5-10\kev$ & $10-20 \kev$   \\
        \hline

        &LFN & 2$\Delta$ (Hz)  & $0.1_{-0.04}^{+0.05}$ & $0.1_{-0.04}^{+0.05}$ & $0.1_{-0.05}^{+0.09}$ \\
        &    &  Norm          & $2.8^{+0.7}_{-0.7} \times 10^{-2}$ & $2.6^{+0.7}_{-0.7}\times 10^{-2}$ & $4.0^{+1.2}_{-1.2}\times 10^{-2}$ \\

 	92016-01-02-01 & HFN &2$\Delta$ (Hz) & $1.4_{-0.6}^{+1.4}$ & $0.9_{-0.2}^{+0.4}$ & $1.6_{-0.6}^{+2.0}$ \\
        (24 sep 2006)  &     & Norm & $2.5^{+0.7}_{-0.8} \times 10^{-2}$ & $3.5^{+0.6}_{-0.7}\times 10^{-2}$ & $4.3^{+0.9}_{-1.0}\times 10^{-2}$ \\
        &&Integrated rms (in percentage)&$22.3^{+1.0}_{-1.0}$&$23.9^{+0.7}_{-0.7}$&$28.0^{+1.0}_{-0.9}$\\

        && $\chi^2/dof$ & 155.5/136 & 105.8/112 & 130.5/136   \\
            
        \hline 

        &LFN & 2$\Delta$ (Hz)  & $0.1_{-0.03}^{+0.04}$ & $0.1_{-0.03}^{+0.03}$ & $0.1_{-0.03}^{+0.04}$ \\
        &    &  Norm          & $2.3^{+0.6}_{-0.6} \times 10^{-2}$ & $3.0^{+0.6}_{-0.6}\times 10^{-2}$ & $3.6^{+0.8}_{-0.8}\times 10^{-2}$ \\

 	92016-01-02-02 & HFN &2$\Delta$ (Hz) & $0.8_{-0.2}^{+0.3}$ & $1.0_{-0.2}^{+0.3}$ & $0.9_{-0.2}^{+0.3}$ \\
        (25 sep 2006)  &     & Norm & $2.7^{+0.5}_{-0.5} \times 10^{-2}$ & $2.8^{+0.5}_{-0.5}\times 10^{-2}$ & $3.7^{+0.7}_{-0.7}\times 10^{-2}$ \\
        &&Integrated rms (in percentage)&$21.5^{+0.4}_{-0.7}$&$23.3^{+0.4}_{-0.4}$&$26.3^{+0.5}_{-0.5}$\\

        && $\chi^2/dof$ & 107.0/112 & 157.3/136 & 180.4/159   \\
            
        \hline 
      \end{tabular}
    \end{center}
    
  \end{table*}

  \begin{figure*}
    \centering \vspace{0cm}

    \includegraphics[scale=0.21,angle=-90]{RXTE_4_11.ps}
    \includegraphics[scale=0.21,angle=-90]{RXTE_11_23.ps}
    \includegraphics[scale=0.21,angle=-90]{RXTE_23_47.ps}
    \caption{ Power density spectra of {\it RXTE} obsID 92016-01-01-00
      (20th September) in energy bands $2-5$ keV, $5-10$ keV and
      $10-20$ keV fitted with lorentzians (top panel of plots) and
      residual (bottom panel of plots).  }\label{fig.xte.pds}
  \end{figure*}

\begin{figure}
  \centering
  \includegraphics[width=20cm,height=12cm,keepaspectratio]{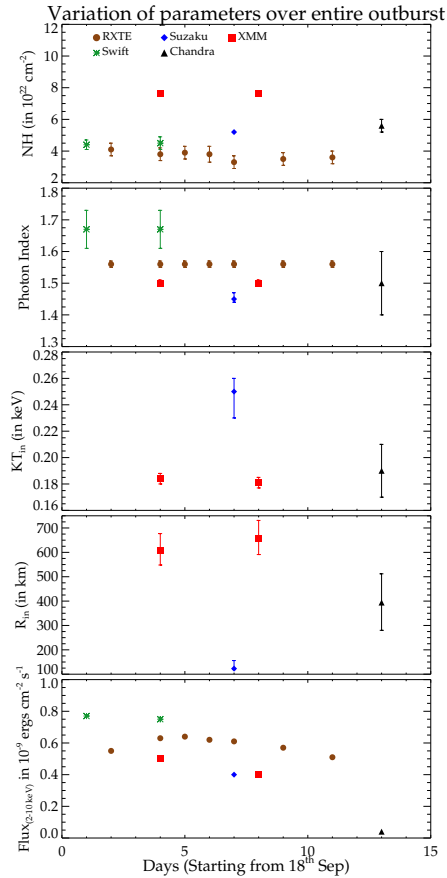}
  \caption{From top to bottom: $N_{H}$, Photon Index, $kT_{in}$,
    $R_{in}$ and Flux$_{2-10\ keV}$ respectively from spectral fits of
    data of different X-ray telescopes (marked in different
    colours). The parameter values plotted have been taken from
    following references: 1. \citet{Rodriguez} for {\it RXTE} (brown),
    absorbed powerlaw is used to calculate $\Gamma$ and model
    {\scshape phabs$\times$edge$\times$comptt} is used to calculate $
    N_{H}$ and Flux.  2. \citet{Walter} for {\it Integral} and {\it
      Swift} (Green), absorbed cutoff powerlaw has been used to
    calculate parameters.  3. our work for {XMM-Newton} (red),
    {\scshape const$\times$tbabs$\times$(nthcomp+diskbb+diskline)} is
    used to calculate parameters. 4. \citet{Paizis} for {\it Suzaku}
    (blue), model {\scshape wabs$\times$edge$\times$cutoffpl} is used
    to calculate $\Gamma$ and {\scshape
      wabs$\times$(gauss+diskbb+compps)} is used to calculate
    $KT_{in}$, $ N_{H}$, $R_{in}$ and Flux. and 5. \citet{Paizis07}
    for {\it Chandra} telescope, {\scshape
      phabs$\times$(diskbb+powerlaw+gauss)} is used to calculate the
    parameters. The flux values in the lower panel are the scaled
    fluxes in the $2-10 \kev$ range derived from broadband model fits
    to respective data sets.}\label{param_var}
\end{figure}

\begin{figure}
  \centering
  \includegraphics[width=7cm,height=7cm,keepaspectratio]{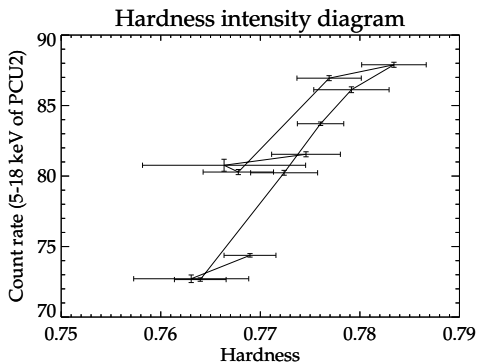}
  \caption{Hardness intensity diagram with X-axis representing the
    hardness of the source and Y-axis the PCU2 counts for $5-18 \kev$
    energy range. The hardness is defined as the ratio of the counts
    in the energy band $8.6-18 \kev$ to $5.0-8.6 \kev$, following
    \citet{2006ARA&A..44...49R} }\label{fig.hid}
\end{figure}

\begin{figure}
  \centering
  \includegraphics[width=7cm,height=7cm,keepaspectratio,angle=-90]{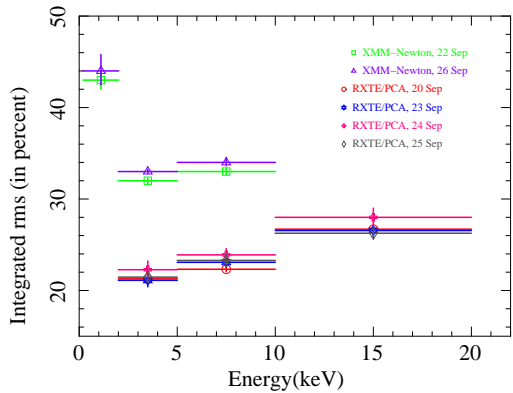}
  \caption{ Variation of integrated rms with different energy bands
    for {\it{XMM-NEWTON/EPIC-pn}} and {\it{RXTE/PCA}}
    observations. Frequency range 0.01-40 Hz is used to calculate
    rms. The rms values are of 68$\%$ confidence limit.}\label{rmsvar}
\end{figure}

\section{Discussion}\label{sec.discuss}
In this work we have presented the spectral and timing analysis of the
two {\it XMM-Newton} observations on 22nd (obs 1) and 26th (obs 2)
September, 2006 of the black hole X-ray binary
IGR~J17497-2821. Following the detection of the source as an X-ray
nova on 2006 17th September, several X-ray telescopes ({\it RXTE,
  Swift, Integral, Suzaku} and {\it Chandra}) have followed the
outburst. The two {\it XMM-Newton} observations were performed during
the middle of the outburst of IGR~J17497-2821 (See
Fig.~\ref{fig.pcu2}).

From our spectral analysis (Sec.~\ref{sec.xmmspectra}), we find that
the $1.2-10\kev$ spectra of IGR~J17497-2821 from both the observations
are well described by a multi-coloured disk blackbody with an inner
disk temperature $kT_{in}\sim 0.2 \kev$ and {\scshape NTHCOMP}
continuum model. The absorbed flux is $f_X = 4.9\times10^{-10}$ ergs
s${}^{-1}$ cm${}^{-2}$ (EPIC-pn) in the $1.2-10\kev$ and corresponding
absorption corrected fluxes are $f_X = 8.6\times10^{-10}$ ergs
s${}^{-1}$ cm${}^{-2}$ (EPIC-pn), for obs 1 (model 4 in
Table~\ref{tab2.fitparams}). The distance to the source is not
known. Considering its close angular proximity to the Galactic centre,
we assume a distance of $\sim 8\kpc$. This gives the observed
luminosity as $\sim 3.5\times10^{36}{\rm~erg~s^{-1}}$ and absorption
corrected luminosity as $ \sim 6.2\times10^{36}{\rm~erg\ s^{-1}}$
(EPIC-pn) in the $1.2-10\kev$ band for obs 1 (model 4 in
Table~\ref{tab2.fitparams}). IGR~J17497-2821 was slightly fainter in
the second observation with an absorbed flux
$f_X=4.5\times10^{-10}{\rm~ergs~cm^{-2}~s^{-1}}$ and corresponding
unabsorbed flux $f_X=8.3\times10^{-10}{\rm~ergs~cm^{-2}~s^{-1}}$ in
the $1.2-10\kev$ band. From the combined analysis of the EPIC-pn and
MOS2 data we find the flux of the disk blackbody component to be $\sim
5$ times lower than that of the power law component in the $1.2-10
\kev$ range. The strong and hard ($\Gamma\sim1.50$) (EPIC-pn) powerlaw
and a weak disk component are typical of black hole X-ray binaries in
the hard state \citep[see
e.g.,][]{2002A&A...390..199B,2004MNRAS.355.1105F,2006ARA&A..44...49R}.

We also detected a broad (FWHM $\sim 27000$\kms) fluorescent Iron
K$\alpha$ line from IGR~J17497--2821 in both the \xmm{}
observations. The apparent inner disk radius $R_{in}$ is obtained from
the normalisation of the {\sc diskbb} component,
\begin{equation}
  N_{DBB}=\left(\frac{R_{in}/{\rm~km}}{D/10 \mbox{ kpc}} \right)^{2}\cos \theta .
\end{equation}
We take inclination and normalization from {\scshape
    diskline} and {\scshape diskbb} component, respectively, and
estimate the limits on the disc inner radius as $400 <
  R_{\rm in} < 703$ km (obs 1) ( model 4 in
    Table~\ref{tab2.fitparams}) and $445 < R_{in} < 613$ km (obs 2)
( model 4 in Table~\ref{tab.fitparams}). The apparent
inner disk radius $R_{in}$ inferred from the {\scshape diskbb}
component is related to the realistic inner disk radius $r_{in}$ by
\begin{equation}
  r_{in} = \xi \kappa^2 R_{in} ,
\end{equation}
\citep{1998PASJ...50..667K} where $\kappa \sim 1.7-2.0$
\citep[e.g.,][]{1995ApJ...445..780S} is the ratio of colour
temperature to effective temperature, and $\xi$ is a correction factor
to account for the fact that the maximum disk temperature $T_{in}$,
occurs at a radius slightly larger than the actual inner radius
$r_{in}$.  \citet{1998PASJ...50..667K} calculated
$\xi=0.412$. However, we note that exclusion of the low energy data
below $\sim 1.2$ keV in the present work (as mentioned in
Sec.~\ref{sec.xmmspectra}) introduces uncertainties in the values of
the disk temperature and radius computed in this work.

The results from the {\it XMM-Newton} spectral analysis are in keeping
with the general trends of the different physical parameters as
observed by various X-ray instruments.  In Fig.~\ref{param_var} we
present the comparison of our results with previous works by
\citet{Walter} (hereafter RW07) on joint spectral fit of {\it
  Swift/XRT} and {\it INTEGRAL} data, analysis of {\it Suzaku} data by
\citet{Paizis} (hereafter AP09), \citet{Rodriguez} (JR07) for {\it
  RXTE} results and \citet{Paizis07} (hereafter AP07) for analysis of
{\it Chandra} data.  It is difficult to directly compare the different
spectral parameters from previous works on this source owing to the
different spectral models used for the analysis. However, qualitative
trends in the variation of different parameters can be ascertained.
The neutral absorption derived from our results ($N_H\sim
7.5\times10^{22}{\rm~cm^{-2}}$) is different from other observations.

The photon index ($\Gamma\sim 1.50$) from the analysis of both
observations is consistent with previous estimates from {\it Suzaku}
data by AP09. However, the $\Gamma$ inferred from the {\it RXTE} and
{\it Swift}-{\it INTEGRAL} observations by JR07 and RW07 during the
outburst is significantly different. AP09 ascribes the difference to
effects from instrumental cross-calibration. Also, the source being
close to the Galactic centre, possible contamination from the Galactic
background may affect spectral analysis of data from {\it RXTE}/PCA
with wide field of view. Indeed the neutron star Swift~J1749.4--2807
(within $\sim 13{\rm~arcmin}$ radius of IGR~J17497--2821) was also
active during this period as reported by \citet{wijnands09}. However,
the qualitative trend for the results from different instruments is
similar, and the $\Gamma$ was nearly constant during the outburst.

{\textit{Suzaku}} and {\it{Chandra}} observations showed the evidence
of an accretion disk with inner disk temperature $kT_{in}\sim0.25\kev$
and $kT_{in}\sim0.2\kev$, respectively, as reported in AP09 and AP07
respectively. This is consistent with the value obtained for our
analysis of the two {\it XMM-Newton} observations ($kT_{in} \sim
0.2\kev$). To compare with previous results, we have computed the
apparent inner disk radius assuming an inclination angle of $60^\circ$
and distance $8\kpc$. We found $R_{in} = 649_{-84}^{+100}{\rm~km}$ and
$715_{-85}^{+108}{\rm~km}$ for obs 1 and obs 2, respectively.
Comparing the results we find that the radii estimated from the {\it
  XMM-Newton} and {\it Chandra} observations are close in value,
within errors, whereas that from the {\it Suzaku} data on 25th,
September is lower than the above estimates, indicating the formation
of an accretion disk. However, the discrepancy could also be be due to
different models and energy bands used for calculation of disk radius
with different instruments. The {\it RXTE/PCA} light curve
(Fig.~\ref{fig.pcu2}) shows a monotonic variation in the count rate.
Fig.~\ref{param_var} shows that the $2-10\kev$ flux measured with
\suzaku{} is significantly lower by $\sim 50\%$ than that measured
with \rxte{}/PCA on the same day. This difference is likely to be due
to the uncertainties in the absolute flux calibration of \rxte{}/PCA
and the burst and window modes of \suzaku{}/XISs. The lower inner
radius measured with the \suzaku{} data may be due to the lower {\sc
  diskbb} normalization which in turn may be related to the
uncertainty in the absolute flux calibration.

The lowest panel of Fig.~\ref{param_var} shows the variation of the
flux scaled to the $2-10 \kev$ energy band, calculated from the
reported best fit model parameters by RW07, JR07, AP07 and AP09. Flux
of the source shows gradual decrease with time with the flux on 2006
October 01 (AP07) being nearly 20 times lower than that at the
beginning on 2006 September 19. We have presented the variation of
hardness ratio (ratio of counts in $8.6-18 \kev$ to $5.0-8.6 \kev$
energy bands) as defined in \citep{2006ARA&A..44...49R}) with the
average PCU2 counts ($5-18 \kev$) in Fig.~\ref{fig.hid}.  The source
is clearly seen to have hardness $\geq 0.76$ throughout the outburst.
From the overall results of the spectral analysis, the source appears
to stay in the hard state during its entire outburst with $\Gamma \leq
2$.

Although differing in frequency and energy range, the full band
($0.3-10\kev$) PDS derived from the timing mode EPIC-pn data are
similar to that obtained from the {\it RXTE} data performed by
\citep{Rodriguez} (JR07).  As in JR07, we require three Lorentzian
components to fit the PDS (Low frequency noise component (LFN),
mid-frequency noise component (MFN) and high frequency noise component
(HFN) for $\nu \geq 1$ Hz). Such broad features is typical of black
hole binaries (e.g.  Cyg x-1 and
GX~339--4~\citep{Nowak}). Characteristic frequencies of LFN, HFN and
MFN are consistent with the
JR07.
For same frequency (0.01-40Hz) and energy (2-10keV) range, the fit
parameters of the LFN, MFN and HFN are similar for both the RXTE and
XMM observations. However the integrated rms for the XMM data was
found to be $\sim 32\%$, larger than that of the RXTE ($\sim 23\%$).

Based on our energy dependent timing analysis of {\it{XMM-Newton}} and
{\it RXTE}, we find that the timing behaviour of IGR~J17497--2821
strongly depends on the energy band. The MFN feature is not present in
the low energy band (0.3-2 keV (XMM-Newton), 2-5 keV (RXTE)), which
could be due to poor signal to noise ratio in this band But variation
of rms is found to be strongly dependent on energy bands. Softer
energy bands had higher fractional rms e.g. rms of the energy band
$0.3-2\kev$ of \xmm{} data was nearly $\sim 1.3$ times that of the
energy bands 2-5 and 5-10 keV. However, at higher energies, the rms
variability was found to to increase with energy. Such a behaviour
though typical for blackholes in high/soft state \citep{Belloni1997},
is not common for sources in hard state. Increased variability at
higher energy in the hard state of some sources e.g.  1E~1740.7-2942
and Cyg X-1 have been reported by \citet{lin2000}.

Also, we see from Fig.~\ref{rmsvar}, that the magnitude
  of the rms variability is lower for {\it RXTE} observations than
  that of {\it XMM} data.  A possible reason for the above could be
  increase of the mean flux due to contribution from the galactic
  ridge emission to the {\it RXTE} data (see Sec.~\ref{ref.append}).
Although the accreting milli-second pulsar Swift~J1749.4--2807, within
the RXTE field of view of IGR~J17497--2821, was also present in
outburst at the same time \citep{wijna09}, its flux was $\sim$ 3000
times less than that of IGR J17497-2821, could not have contributed to
the background.

\section{Acknowledgement}
The authors would like to extend sincere thanks to the referee for
his/her critical comments which has contributed considerbaly in
improving the work. The authors would also like to thank the
organizing committee of the 2nd X-ray Astronomy School at IUCAA during
Feb 4 -- March 2, 2013 where this work was initiated. Md. Shah Alam
and Aditya S. Mondal would also like to thank IUCAA for hosting them
during subsequent visits. SJ and GCD acknowledge support from grant
under ISRO-RESPOND programme(ISRO/RES/2/384/2014-15).  Md. Shah Alam
would like to acknowledge fellowship provided under ISRO-RESPOND
programme .  BR likes to thank IUCAA for the hospitality and
facilities extended to him under their Visiting Associateship
programme.  This research has made use of the General High-energy
Aperiodic Timing Software (GHATS) package developed by Dr. Tomaso
Belloni at INAF - Osservatorio Astronomico di Brera. We would also
like to thank Dr. Dipankar Bhattacharya, Dr. Ranjeev Misra and
Dr. Tomaso Belloni for helpful suggestions and discussions.

\appendix \label{ref.append}
\section{Diminished rms variability due to contamination from the
  Galactic ridge emission}  \rxte{}/PCA being a
  non-imaging detector with a large field of view, is more susceptible
  to contamination from background X-ray flux due to the Galactic
  ridge emission, as opposed to {\it XMM-Newton}. Excess background
  can lower the net observed fractional rms variability
  (rms/mean). Following \citet{Rodriguez} if we consider the \rxte{}
  obsID 30185-01-20-00\footnote{Observed at the end of the 1998
    outburst of XTE~J1748-288} to represent the Galactic background
  emission, the mean count rate of the background emission then was
  found to be $\sim 30$ counts~s$^{-1}$ in the 2-10 keV band. Using
  the WebPIMMS
  tool\footnote{\url{http://heasarc.gsfc.nasa.gov/cgi-bin/Tools/w3pimms/w3pimms.pl}},
  we find the equivalent RXTE/PCA count rate of the 2-10~keV flux of
  IGR J17497-2821 measured with XMM-Newton to be 51 counts
  s$^{-1}$. Thus the sum of the estimated \rxte{}/PCA count rate and
  the Galactic background is $\sim 81$ counts s$^{-1}$. This is
  similar to the mean 2-10~keV count rate of 80 counts/s measured from
  the \rxte{} obsID 92016-01-00. Thus it is entirely possible that the
  addition of background flux due to the Galactic ridge emission to
  the source flux can lower the observed rms variability of
  IGR~J17497--2821, as seen in our case.

\label{lastpage}    
\def\apj{ApJ}%
\def\araa{ARAA}%
\def\mnras{MNRAS}%
\def\pasp{PASP}%
\def\aap{A\&A}%
\def\apjl{ApJ} \def\physrep{PhR} \def\apjs{ApJS} \def\pasa{PASA}
\def\pasj{PASJ} \def\nat{Nature} \def\memsai{MmSAI}

\bibliographystyle{mn2e}

\end{document}